\documentclass{article}
\usepackage{spconf,amsmath,graphicx,hyperref}
\usepackage{caption}

\usepackage{booktabs}

\title{Query-Conditioned Liquid Temporal Grounding for Long-Form Audio}
\name{
Yuelan Cheng\textsuperscript{1},
Yonggang Zhu\textsuperscript{3},
Yongqiang Chen\textsuperscript{1},
Peiwei Chang\textsuperscript{1},
Yingyi Chen\textsuperscript{2},
Wenwu Wang\textsuperscript{1}
}
\address{
\textsuperscript{1}Centre for Vision, Speech and Signal Processing, University of Surrey, UK\\
\textsuperscript{2}College of Information and Electrical Engineering, China Agricultural University, China\\
\textsuperscript{3}School of Artificial Intelligence, Beijing University of Posts and Telecommunications, China
}
\begin{document}
%
\maketitle

\begin{abstract}
Long-form temporal audio grounding (TAG) is challenging because target events often occupy only a small portion of a recording, while queries may impose multiple semantic constraints over a large temporal search space. We propose a query-conditioned liquid temporal grounding framework that injects query semantics through FiLM before a closed-form
continuous-time (CfC) network models long-range temporal dependencies. Multiple query-relevant candidates are generated, from which a frozen large audio-language model selects the most compatible one for lightweight local boundary refinement. The pretrained backbone remains frozen, with only lightweight grounding modules optimized. Experiments on the English TAG subset of LAT-Bench show consistent improvements across MOSS-Audio and Qwen3-Omni, while ablations confirm the complementary roles of FiLM and CfC.
\end{abstract}

%
\vspace{-10pt}

\section{Introduction}
\label{sec:intro}

\begin{figure}[!t]
    \centering
    \vspace{-10pt}
    \includegraphics[width=0.40\textwidth]{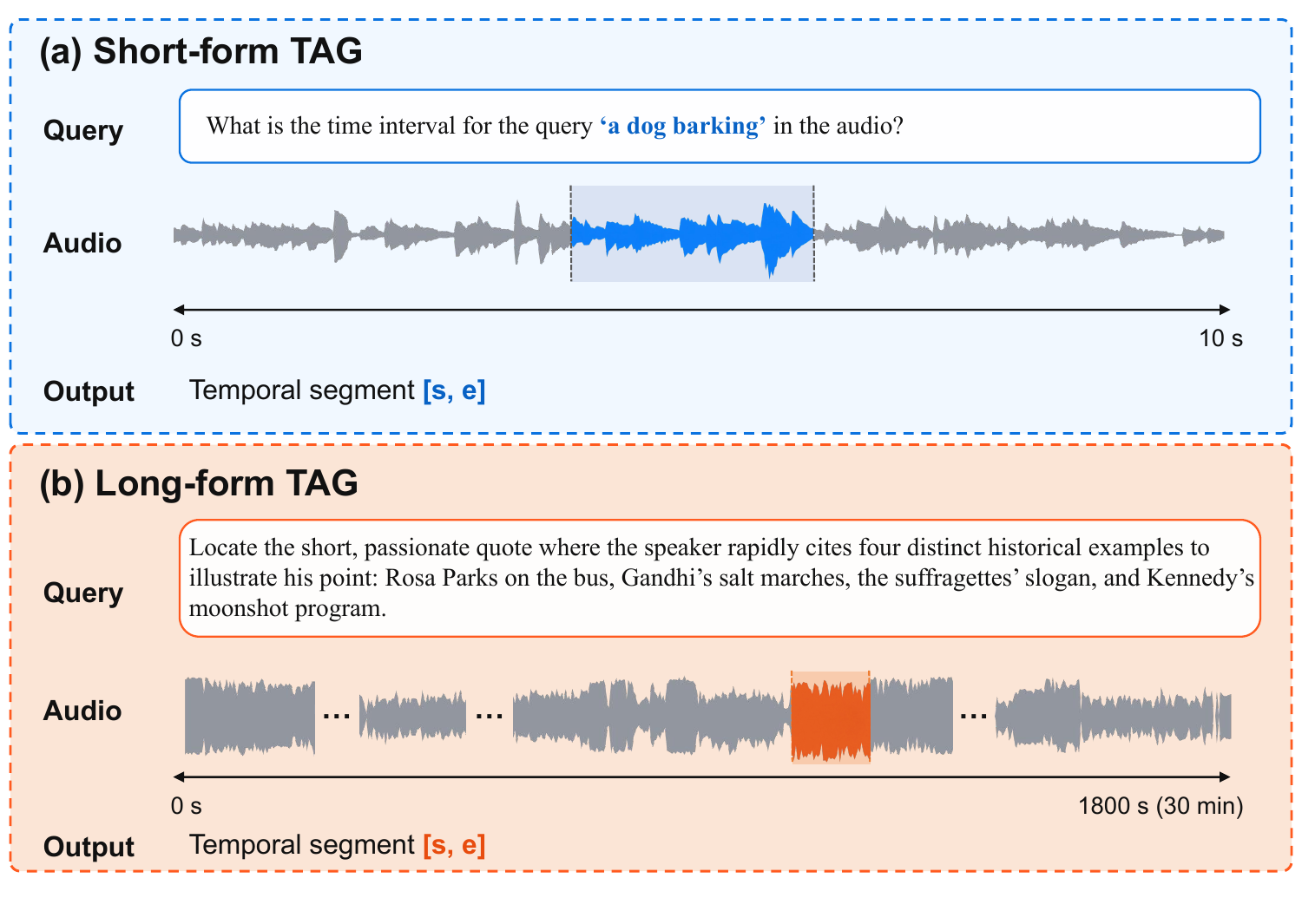}
    \caption{Comparison between short-form and long-form temporal audio
    grounding.}
    \vspace{-10pt}
    \label{fig:tag_comparison}
\end{figure}

\looseness=-1 Temporal audio grounding (TAG) aims to localize the temporal boundaries of audio content described by a natural-language query~\cite{xu2021text,xu2024towards}. Early TAG studies focus on learning fine-grained correspondences between textual descriptions and localized acoustic events, while recent
audio-language pretraining methods further improve fine-grained audio-text alignment~\cite{li2024advancing,li2026finelap}. Long-form TAG extends temporal grounding to substantially longer recordings, introducing greater localization difficulty. The target segment may occupy only a small fraction of the recording and can be surrounded by acoustically or semantically similar content~\cite{shao2026listening,dai2026tag}. Meanwhile, the grounding query may contain multiple semantic constraints involving speakers, events, attributes, and temporal relations~\cite{shao2026listening}. As illustrated in Fig.~\ref{fig:tag_comparison}, long-form TAG requires localizing short target segments corresponding to potentially complex queries over a substantially larger temporal search space. Recent studies also report substantial performance degradation as audio duration increases and show that precise temporal localization remains challenging for current audio-language models~\cite{wang2026timeaudio,shao2026listening,dai2026tag}.

Existing approaches improve temporal audio understanding through explicit
temporal encoding, grounding supervision, or global-to-local
reasoning~\cite{wang2026timeaudio,sun2026spotsound,shao2026listening}.
However, long-form TAG requires searching for query-relevant events over
a large temporal space, motivating temporal modeling that is directly
guided by the query. To this end, we propose a query-conditioned liquid
temporal grounding framework. FiLM~\cite{perez2018film} first injects
query semantics into temporal audio representations, which are then
modeled by a closed-form continuous-time (CfC) network~\cite{hasani2022closed}.
Here, ``liquid'' refers to the input-adaptive continuous-time dynamics of
CfC, allowing temporal evolution to depend on the query-conditioned
representations. The resulting temporal evidence is used to generate
multiple candidate intervals, from which a frozen LALM selects the most
semantically compatible one. A lightweight local refiner further adjusts
its boundaries to obtain the final prediction. The pretrained
audio-language backbone remains frozen throughout training, and only the
proposed temporal grounding modules are optimized, enabling training on
a single NVIDIA A100 GPU while maintaining competitive localization
performance.

\begin{figure*}[!t]
    \centering
    \vspace{-10pt}
\includegraphics[width=0.8\textwidth]{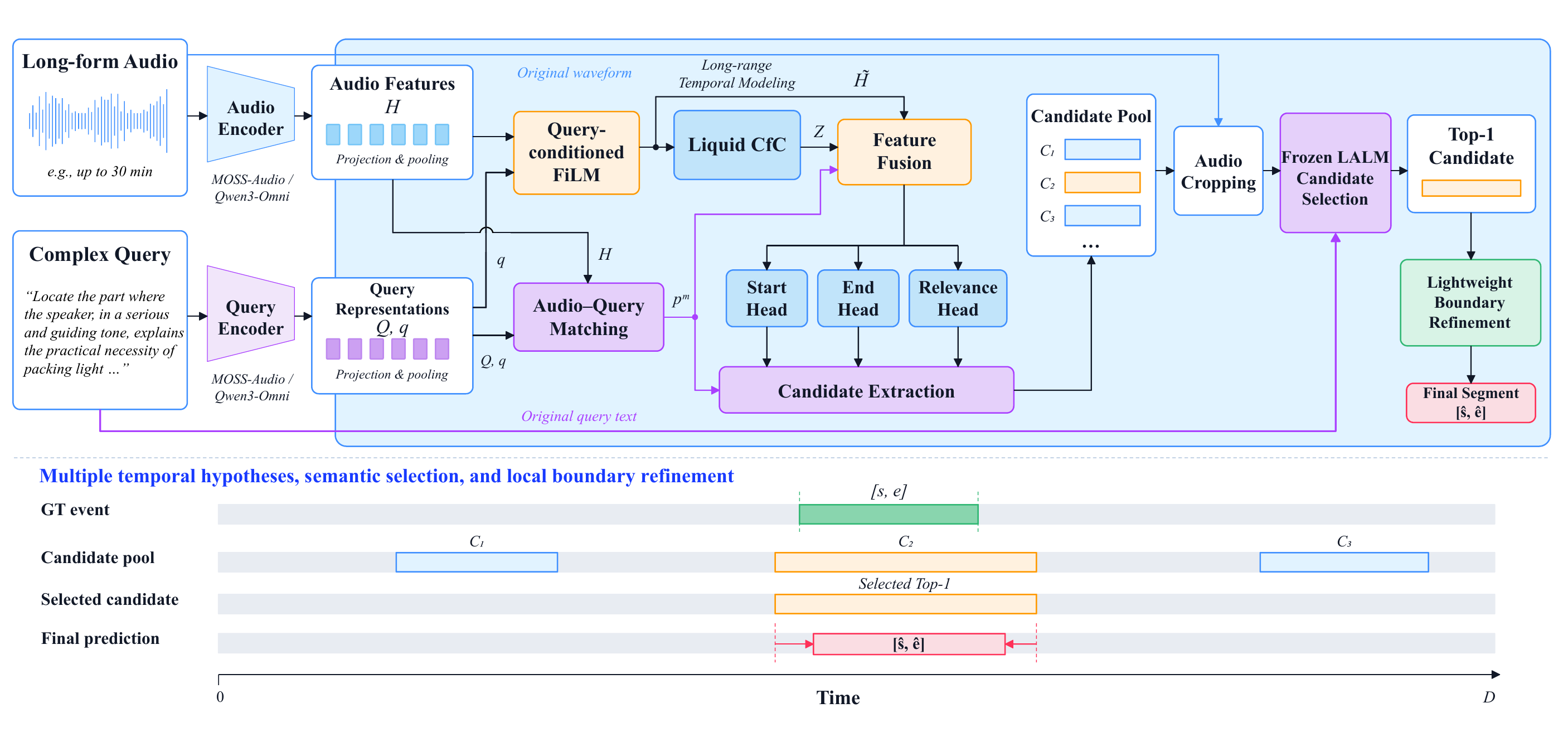}
\vspace{-12pt}
    \caption{Overview of the proposed query-conditioned liquid temporal grounding framework.}
    \label{fig:method}
    \vspace{-7pt}
\end{figure*}

Experiments on the English TAG subset of LAT-Bench~\cite{shao2026listening} demonstrate consistent improvements with both MOSS-Audio~\cite{yang2026moss} and Qwen3-Omni~\cite{xu2025qwen3} backbones. Ablation studies demonstrate a strong interaction between FiLM-based query conditioning and CfC-based temporal modeling. Lightweight boundary refinement provides additional gains in localization precision.

\vspace{-3pt}
\section{Proposed Method}
\label{sec:method}
\vspace{-7pt}

\subsection{Framework Overview}
\vspace{-5pt}

As illustrated in Fig.~\ref{fig:method}, our framework performs query-conditioned temporal grounding over long-form audio $\mathcal{A}$. Instead of directly regressing a single temporal interval from the full recording, the model first preserves multiple query-relevant temporal hypotheses through query-conditioned temporal modeling. A frozen large audio-language model (LALM)~\cite{xu2025qwen3,yang2026moss} then selects the candidate that is most semantically compatible with the query, followed by lightweight local boundary refinement to obtain the final prediction $\hat{\tau}=(\hat{s},\hat{e})$.

\vspace{-8pt}
\subsection{Query-Conditioned Temporal Modeling}
\vspace{-5pt}

The audio encoder first produces high-resolution audio representations, which are temporally projected and pooled into
$\mathbf{H}=[\mathbf{h}_1,\ldots,\mathbf{h}_T]$.
The query encoder produces token-level representations
$\mathbf{Q}=[\mathbf{q}_1,\ldots,\mathbf{q}_L]$
and a pooled query representation $\mathbf{q}$.

Since $\mathbf{H}$ is initially query-independent, we first inject query semantics using feature-wise linear modulation (FiLM)~\cite{perez2018film}:
\begin{equation}
\tilde{\mathbf{h}}_t
=
\left(1+f_{\gamma}(\mathbf{q})\right)
\odot\mathbf{h}*t
+f*{\beta}(\mathbf{q}),
\end{equation}
where $f_{\gamma}$ and $f_{\beta}$ are learnable projections. The same query-conditioned modulation parameters are shared across all temporal steps, allowing the model to emphasize query-relevant feature dimensions while preserving temporal order.

The query-conditioned sequence $\tilde{\mathbf{H}}=[\tilde{\mathbf{h}}_1,\ldots,\tilde{\mathbf{h}}_T]$
is then processed by a closed-form continuous-time (CfC) network~\cite{hasani2022closed}:
\begin{equation}
\mathbf{x}_t=\mathrm{CfCCell}(\tilde{\mathbf{h}}_t,\mathbf{x}_{t-1};\Delta_t).
\end{equation}
The resulting states are normalized and regularized as
$\mathbf{z}_t=\mathrm{Dropout}(\mathrm{LN}(\mathbf{x}_t))$, yielding
$\mathbf{Z}=[\mathbf{z}_1,\ldots,\mathbf{z}_T]$, where
$\Delta_t=D/(T-1)$ for a recording of duration $D$.
Thus, the CfC dynamics depend on both the query-conditioned audio representation and the physical time span represented by each temporal step. Since FiLM is applied before CfC, query semantics directly influence
long-range temporal information propagation rather than only downstream candidate scoring.

\vspace{-8pt}
\subsection{Temporal Candidate Generation}
\vspace{-5pt}

In parallel with the query-conditioned temporal modeling branch, we compute a frame-wise audio--query matching signal $p^m$ from the original temporal representations $\mathbf{H}$. Audio frames interact with the query tokens to obtain a token-level matching score $S_i^{\mathrm{tok}}$, which is combined with the global similarity $S_i^{\mathrm{glb}}$ to the pooled query representation:
\begin{equation}
p_i^m
=
(1-w_g)S_i^{\mathrm{tok}}
+
w_gS_i^{\mathrm{glb}}.
\end{equation}

The matching signal is then fused with the FiLM-conditioned features and CfC outputs as
$\mathbf{F}=\mathrm{LN}([\tilde{\mathbf{H}};\mathbf{Z};p^m])$,
from which three lightweight heads predict the start, end, and frame-wise relevance distributions
$p^s$, $p^e$, and $p^r$, respectively.

Candidate intervals are generated from high-response regions of the relevance and matching signals. We first combine the normalized relevance and matching curves:
\begin{equation}
u=w_r\bar{p}^{r}+w_m\bar{p}^{m}.
\end{equation}
Seeds are selected above a fixed seed quantile $\rho_s=0.85$ of $u$
and grown toward neighboring positions exceeding a growth quantile
$\rho_g$, where $\rho_g=0.65$ by default.
Nearby regions are then merged, and multi-scale temporal variants are generated.
Finally, local responses in $p^s$ and $p^e$ are used to adjust the proposal
boundaries, yielding the candidate set
$\mathcal{C}=\{c_i=(s_i,e_i)\}_{i=1}^{M}$.
The candidate extraction procedure itself contains no trainable parameters
and operates on the temporal evidence predicted by the model.

\noindent\textbf{Training objective.}
The temporal predictor is trained to estimate boundary locations,
identify relevant regions, and learn audio--query correspondence.
The boundary loss $\mathcal{L}_{\mathrm{bnd}}$ supervises the
start and end distributions using soft boundary targets.
A focal-style class-balanced binary cross-entropy loss
$\mathcal{L}_{\mathrm{rel}}$ supervises frame-wise relevance
to address the foreground--background imbalance in long recordings.
For the matching signal $p^m$, a hard-negative loss
$\mathcal{L}_{\mathrm{match}}$ encourages responses inside
the ground-truth interval to exceed strong responses outside it.
We additionally apply a one-dimensional GIoU loss
$\mathcal{L}_{\mathrm{gIoU}}$ to supervise the predicted interval
and an end-boundary hard-negative loss
$\mathcal{L}_{\mathrm{ehn}}$ to suppress spurious end responses
far from the ground-truth endpoint.

The overall training objective is
\begin{equation}
\begin{aligned}
\mathcal{L}_{\mathrm{gen}}
={}&
\lambda_{\mathrm{bnd}}\mathcal{L}_{\mathrm{bnd}}
+
\lambda_{\mathrm{rel}}\mathcal{L}_{\mathrm{rel}}
+
\lambda_{\mathrm{match}}\mathcal{L}_{\mathrm{match}}
\\
&+
\lambda_{\mathrm{gIoU}}\mathcal{L}_{\mathrm{gIoU}}
+
\lambda_{\mathrm{ehn}}\mathcal{L}_{\mathrm{ehn}},
\end{aligned}
\label{eq:gen_loss}
\end{equation}
where the $\lambda$ coefficients control the contributions
of the individual losses.
These losses optimize the temporal evidence predictor,
while candidate extraction remains parameter-free.
They do not directly supervise candidate-level ranking;
semantic candidate selection is subsequently performed
by the frozen LALM.

\vspace{-8pt}
\subsection{LALM-Based Candidate Selection}
\vspace{-5pt}

For each candidate interval $c_i=(s_i,e_i)$, the frozen
LALM independently evaluates the corresponding audio crop
together with the original query $\mathcal{Q}$ using a
Yes/No compatibility prompt. We apply a two-way softmax
to the next-token logits for \emph{Yes} and \emph{No},
and use the probability assigned to \emph{Yes} as the
candidate's compatibility score. The highest-scoring
candidate is selected, yielding the interval
$\tilde{\tau}=(\tilde{s},\tilde{e})$.

\vspace{-8pt}
\subsection{Local Boundary Refinement}
\vspace{-5pt}
For the Qwen3-Omni system, we apply a lightweight local refiner to the
LALM-selected candidate $\tilde{\tau}=(\tilde{s},\tilde{e})$.
A local window extending 5 seconds beyond both candidate boundaries is
encoded through two complementary paths: a frozen Qwen3-Omni audio
encoder provides semantic representations, while a lightweight Conv1D
encoder operating on the mel spectrogram provides higher-resolution
acoustic features. Both representations are query-conditioned by FiLM,
fused, and pooled around the candidate start, center, and end to obtain
a local boundary representation.

The refiner predicts bounded residual corrections and gates,
$\Delta b=\delta_{\max}\tanh(r_b)$ and
$\alpha_b=\sigma(g_b)$, for $b\in\{s,e\}$, where
$\delta_{\max}=5$ seconds. The refined boundaries are computed as
$\hat{b}=\mathrm{clip}(\tilde{b}+\alpha_b\Delta b,0,D)$, yielding the
final interval $\hat{\tau}=(\hat{s},\hat{e})$.

The refiner is trained with
$\mathcal{L}_{\mathrm{ref}}
=\lambda_{\mathrm{reg}}\mathcal{L}_{\mathrm{reg}}
+\lambda_{\mathrm{IoU}}\mathcal{L}_{\mathrm{IoU}}$,
where $\mathcal{L}_{\mathrm{reg}}$ is a SmoothL1 loss on the bounded
pre-gate boundary residuals and
$\mathcal{L}_{\mathrm{IoU}}
=1-\operatorname{IoU}([\hat{s},\hat{e}],[s^{*},e^{*}])$.
Candidates requiring a start or end correction larger than
$\delta_{\max}$ are excluded during training, since such targets fall
outside the correction range of the local refiner.


\section{EXPERIMENTS}
\label{sec:pagestyle}
\vspace{-10pt}
\noindent\textbf{Datasets.}
We conduct experiments on the English TAG data from
\textit{Listening with Time}~\cite{shao2026listening}.
For training, we use the English TAG subset of LAT Chronicle, containing 3,737 query--segment pairs, with a 9:1 train--validation split. Evaluation is conducted on the English TAG subset of LAT-Bench, which contains long-form recordings of up to 30 minutes with corresponding grounding queries.

\noindent\textbf{Evaluation Metrics.}
Following the official LAT-Bench protocol, we report mean intersection over union (mIoU) and Recall@IoU at thresholds of 0.3, 0.5, and 0.7.

\noindent\textbf{Implementation Details.}
We use frozen MOSS-Audio-8B-Instruct~\cite{yang2026moss} and
Qwen3-Omni-30B-A3B-Instruct~\cite{xu2025qwen3} as backbones.
The temporal grounding modules are trained for three epochs using AdamW
with a learning rate of $1\times10^{-4}$ over 1,024 temporal bins.
The loss weights in Eq.~\eqref{eq:gen_loss} are set to
$(0.3, 1.0, 0.5, 0.5, 0.3)$.
For Qwen3-Omni, the LALM-selected Top-1 candidate is further processed
by a lightweight boundary refiner trained on a $\pm5$\,s local window.
All experiments are conducted on a single NVIDIA A100 GPU.

\section{Results}
\label{sec:results}
\vspace{-5pt}

\subsection{Main Localization Results}
\label{sec:main_results}
\vspace{-5pt}

We evaluate the proposed framework with two audio-language backbones,
MOSS-Audio and Qwen3-Omni. For reference, we also include the official
English TAG results reported in \textit{Listening with Time}
~\cite{shao2026listening}. The results are summarized in
Table~\ref{tab:main_results}.

\begin{table}[t]
\centering
\caption{Temporal grounding performance on the English TAG subset of
LAT-Bench. SW denotes sliding-window localization, QA-SFT denotes
QA-only supervised fine-tuning, and Ref. denotes boundary refinement.
Union combines the candidate pools generated with $\rho_g{=}0.65$ and
$\rho_g{=}0.80$ before LALM-based selection. All metrics are percentages.}
\vspace{-5pt}
\label{tab:main_results}
\footnotesize
\setlength{\tabcolsep}{3.5pt}
\renewcommand{\arraystretch}{1.0}

\begin{tabular}{@{}llcccc@{}}
\toprule
\textbf{Model} &
\textbf{Setting} &
\textbf{mIoU} &
\textbf{R@0.3} &
\textbf{R@0.5} &
\textbf{R@0.7} \\
\midrule

\multicolumn{6}{l}{\textit{Reported results~\cite{shao2026listening}}} \\

LAT-Audio
& Main Model
& 50.00 & 68.10 & 54.10 & 34.50 \\

Gemini-2.5-Pro
& Main Model
& 45.30 & 65.20 & 53.90 & 27.70 \\

Gemini-3.0-Pro
& Main Model
& 41.00 & 51.40 & 44.80 & 22.90 \\

Qwen3-Omni
& SW
& 26.20 & 41.90 & 25.80 & 15.50 \\

Qwen3-Omni
& QA-SFT
& 39.20 & 54.20 & 41.70 & 24.50 \\

\midrule
\multicolumn{6}{l}{\textit{Our experiments}} \\

MOSS-Audio
& Direct
& 15.71 & 22.77 & 14.79 & 8.69 \\

MOSS-Audio
& SW
& 26.39 & 36.15 & 14.79 & 4.23 \\

MOSS-Audio
& QA-SFT
& 17.75 & 27.00 & 18.08 & 9.86 \\

MOSS-Audio
& \textbf{Ours}
& \textbf{38.52}
& \textbf{57.98}
& \textbf{40.61}
& \textbf{19.25} \\

\midrule

Qwen3-Omni
& Direct
& 15.80 & 26.40 & 16.00 & 7.00 \\

Qwen3-Omni
& Ours ($\rho_g{=}0.65$)
& 42.72 & 65.02 & 42.49 & 19.95 \\

Qwen3-Omni
& Ours ($\rho_g{=}0.65$) + Ref.
& 43.20 & \textbf{65.26} & 42.96 & 21.13 \\

Qwen3-Omni
& Ours (Union)
& 43.45 & \textbf{65.26} & 45.54 & 21.13 \\

Qwen3-Omni
& \textbf{Ours (Union) + Ref.}
& \textbf{44.13}
& {65.02}
& \textbf{46.95}
& \textbf{22.07} \\

\bottomrule
\end{tabular}
\vspace{-4pt}
\end{table}


The proposed framework consistently improves temporal grounding across
both audio-language backbones. With MOSS-Audio, our method achieves
38.52 mIoU and 40.61 R@0.5, outperforming direct localization,
sliding-window localization, and QA-SFT. On Qwen3-Omni, the
$\rho_g=0.65$ configuration reaches 42.72 mIoU, exceeding the reported
SW and QA-SFT results in mIoU. While LAT-Audio achieves the strongest
absolute localization performance among the reported systems, our method focuses on lightweight adaptation of frozen audio-language backbones and shows consistent gains across different backbone architectures.

Combining the candidate pools generated with $\rho_g{=}0.65$ and $\rho_g{=}0.80$ further improves Qwen3-Omni performance. Compared with the $\rho_g{=}0.65$ configuration, the union pool increases mIoU from 42.72 to 43.45, R@0.5 from 42.49 to 45.54, and R@0.7 from 19.95 to 21.13. This suggests that the two proposal configurations provide complementary
temporal hypotheses, allowing the frozen LALM to select from a more diverse candidate pool.

The lightweight boundary refiner provides additional gains under both
candidate-pool configurations. For $\rho_g{=}0.65$, mIoU improves from
42.72 to 43.20 and R@0.7 from 19.95 to 21.13. For the union pool, mIoU
increases from 43.45 to 44.13, while R@0.5 and R@0.7 increase from
45.54 to 46.95 and from 21.13 to 22.07, respectively. These improvements
across different candidate pools support the effectiveness of local
boundary refinement. Although the refined system remains below QA-SFT
at R@0.7, it achieves higher overall mIoU without end-to-end backbone
adaptation.



\vspace{-10pt}
\subsection{Candidate Pool Analysis}
\label{sec:candidate_analysis}
\vspace{-8pt}

We further analyze the generated candidate pools by comparing the LALM-selected Top-1 predictions with oracle selection before boundary refinement. For each candidate pool, the oracle selects the candidate
with the highest IoU to the ground-truth interval from the same pool. The results are reported in Table~\ref{tab:candidate_results}.

\begin{table}[t]
\centering
\caption{Pre-refinement Top-1 and oracle performance of different
candidate pools. Oracle denotes selection of the candidate with the
highest IoU to the ground-truth interval from the same pool.
All metrics are percentages.}
\vspace{-5pt}
\label{tab:candidate_results}
\renewcommand{\arraystretch}{0.92}
\setlength{\tabcolsep}{3.5pt}
\footnotesize

\begin{tabular}{llccccc}
\toprule
\textbf{Backbone} &
\textbf{Pool} &
\textbf{Selection} &
\textbf{mIoU} &
\textbf{R@0.3} &
\textbf{R@0.5} &
\textbf{R@0.7} \\
\midrule

MOSS-Audio
& Ours
& Top-1
& 38.52 & 57.98 & 40.61 & 19.25 \\

&
& Oracle
& 46.72 & 68.31 & 50.94 & 26.76 \\

\midrule

Qwen3-Omni
& $\rho_g{=}0.65$
& Top-1
& 42.72 & 65.02 & 42.49 & 19.95 \\

&
& Oracle
& 51.04 & 74.41 & 55.63 & 30.52 \\

\addlinespace[1pt]

Qwen3-Omni
& $\rho_g{=}0.80$
& Top-1
& 41.47 & 62.44 & 43.90 & 18.31 \\

&
& Oracle
& 50.66 & 74.18 & 54.93 & 31.22 \\

\addlinespace[1pt]

Qwen3-Omni
& Union
& Top-1
& 43.45 & 65.26 & 45.54 & 21.13 \\

&
& Oracle
& 55.20 & 78.40 & 61.97 & 38.26 \\

\bottomrule
\end{tabular}
\end{table}

The oracle consistently outperforms the LALM-selected Top-1 predictions,
showing that the candidate pools often preserve higher-quality temporal
hypotheses than those finally selected. Combining the two Qwen3-Omni
proposal configurations further improves both Top-1 and oracle performance,
suggesting that the union pool provides more diverse and complementary
temporal hypotheses. The remaining Top-1--oracle gap indicates that candidate
selection remains a key bottleneck, particularly under stricter IoU thresholds
requiring more precise localization.

\vspace{-10pt}
\subsection{Ablation Study}
\label{sec:ablation}

We ablate FiLM and CfC on both audio-language backbones and further replace CfC with a standard GRU to examine the role of liquid temporal modeling. To isolate the effects of these components, all ablation experiments use the same $\rho_g{=}0.65$ candidate-generation configuration.
The results are reported in Table~\ref{tab:global_ablation}.

\begin{table}[t]
\centering
\caption{Ablation of query conditioning and temporal modeling.
All metrics are percentages. Bold indicates the best result
within each backbone.}
\vspace{-5pt}
\label{tab:global_ablation}
\small
\renewcommand{\arraystretch}{1.0}
\setlength{\tabcolsep}{4pt}
\begin{tabular}{ccrrrr}
\toprule
\textbf{FiLM} &
\textbf{Temporal} &
\textbf{mIoU} &
\textbf{R@0.3} &
\textbf{R@0.5} &
\textbf{R@0.7} \\
\midrule

\multicolumn{6}{l}{\textit{MOSS-Audio}} \\
No  & None & 36.29 & 54.92 & 36.38 & 18.31 \\
Yes & None & 33.85 & 51.88 & 34.04 & 17.14 \\
No  & CfC  & 37.21 & 54.67 & 37.79 & \textbf{19.71} \\
Yes & GRU  & 32.00 & 46.95 & 32.86 & 17.84 \\
Yes & CfC  & \textbf{38.52} & \textbf{57.98}
           & \textbf{40.61} & 19.25 \\

\midrule
\multicolumn{6}{l}{\textit{Qwen3-Omni}} \\
No  & None & 35.43 & 53.52 & 37.10 & 15.25 \\
Yes & None & 35.31 & 53.29 & 37.79 & 17.14 \\
No  & CfC  & 38.37 & 59.62 & 39.44 & 14.79 \\
Yes & GRU  & 31.29 & 48.83 & 30.05 & 12.68 \\
Yes & CfC  & \textbf{42.72} & \textbf{65.02}
           & \textbf{42.49} & \textbf{19.95} \\

\bottomrule
\end{tabular}
\vspace{-5pt}
\end{table}

\looseness=-1
The full FiLM+CfC configuration achieves the highest mIoU on both backbones. FiLM alone does not improve mIoU over using neither component, whereas adding FiLM to CfC increases mIoU from 37.21 to 38.52 on
MOSS-Audio and from 38.37 to 42.72 on Qwen3-Omni. This suggests that query conditioning becomes more effective when coupled with CfC-based liquid temporal modeling.

More importantly, replacing CfC with a standard GRU reduces mIoU to 32.00 on MOSS-Audio and 31.29 on Qwen3-Omni. The gain therefore cannot be explained simply by introducing recurrent temporal modeling. Instead, the results are consistent with the benefit of the continuous-time, input-adaptive dynamics provided by CfC for modeling query-conditioned temporal evolution in long-form audio.

\section{CONCLUSION}
\label{sec:print}
 \vspace{-5pt}

We presented a query-conditioned liquid temporal grounding framework for
long-form audio using FiLM, CfC, frozen LALM candidate selection, and
lightweight boundary refinement. Experiments on LAT-Bench show consistent
improvements across MOSS-Audio and Qwen3-Omni, while ablations confirm the
complementary roles of FiLM and CfC. The multi-candidate design further
preserves diverse temporal hypotheses, and local refinement improves boundary
precision after candidate selection. With frozen backbones and lightweight
trainable modules, the framework supports single-GPU training. The Top-1--oracle gap motivates future work on stronger candidate selection.

\vfill\pagebreak
\clearpage

\bibliographystyle{IEEEbib}
\bibliography{strings,refs}

@inproceedings{xu2021text,
  title={Text-to-Audio Grounding: Building Correspondence Between Captions and Sound Events},
  author={Xu, Xuenan and Dinkel, Heinrich and Wu, Mengyue and Yu, Kai},
  booktitle={ICASSP 2021-2021 IEEE international conference on acoustics, speech and signal processing (ICASSP)},
  pages={606--610},
  year={2021},
  organization={IEEE}
}

@article{xu2024towards,
  title={Towards Weakly Supervised Text-to-Audio Grounding},
  author={Xu, Xuenan and Ma, Ziyang and Wu, Mengyue and Yu, Kai},
  journal={IEEE Transactions on Multimedia},
  volume={26},
  pages={11126--11138},
  year={2024},
  publisher={IEEE}
}

@inproceedings{wang2026timeaudio,
  title={Listening Between the Frames: Bridging Temporal Gaps in Large Audio-Language Models},
  author={Wang, Hualei and Li, Yiming and Ma, Shuo and Liu, Hong and Wang, Xiangdong},
  booktitle={Proceedings of the AAAI Conference on Artificial Intelligence},
  volume={40},
  number={31},
  pages={26233--26241},
  year={2026}
}

@article{shao2026listening,
  title={Listening with Time: Precise Temporal Awareness for Long-Form Audio Understanding},
  author={Shao, Mingchen and Su, Hang and Tian, Wenjie and Mu, Bingshen and Lin, Zhennan and Fan, Lichun and Luo, Zhenbo and Luan, Jian and Xie, Lei},
  journal={arXiv preprint arXiv:2604.22245},
  year={2026}
}

@article{dai2026tag,
  title={TAG-Bench: Benchmarking Temporal Audio Grounding in Large Audio Language Models},
  author={Dai, Yuhang and Shu, Xin and Li, Zengxi and Xie, Lei and Li, Xiangang and Yu, Jianwei},
  journal={arXiv preprint arXiv:2609.01542},
  year={2026}
}

@article{sun2026spotsound,
  title={SpotSound: Enhancing Large Audio-Language Models with Fine-Grained Temporal Grounding},
  author={Sun, Luoyi and Zhou, Xiao and Li, Zeqian and Zhang, Ya and Wang, Yanfeng and Xie, Weidi},
  journal={arXiv preprint arXiv:2604.13023},
  year={2026}
}

@article{yang2026moss,
  title={MOSS-Audio Technical Report},
  author={Yang, Chen and Yu, Chufan and Chen, Hanfu and Zhu, Jie and Chen, Jingqi and Chen, Ke and Wang, Wenxuan and Wang, Yang and Jiang, Yaozhou and Jiang, Yi and others},
  journal={arXiv preprint arXiv:2606.01802},
  year={2026}
}

@article{xu2025qwen3,
  title={Qwen3-Omni Technical Report},
  author={Xu, Jin and Guo, Zhifang and Hu, Hangrui and Chu, Yunfei and Wang, Xiong and He, Jinzheng and Wang, Yuxuan and Shi, Xian and He, Ting and Zhu, Xinfa and others},
  journal={arXiv preprint arXiv:2509.17765},
  year={2025}
}

@inproceedings{perez2018film,
  title={FiLM: Visual Reasoning with a General Conditioning Layer},
  author={Perez, Ethan and Strub, Florian and De Vries, Harm and Dumoulin, Vincent and Courville, Aaron},
  booktitle={Proceedings of the AAAI conference on artificial intelligence},
  volume={32},
  number={1},
  year={2018}
}

@article{hasani2022closed,
  title={Closed-form continuous-time neural networks},
  author={Hasani, Ramin and Lechner, Mathias and Amini, Alexander and Liebenwein, Lucas and Ray, Aaron and Tschaikowski, Max and Teschl, Gerald and Rus, Daniela},
  journal={Nature Machine Intelligence},
  volume={4},
  number={11},
  pages={992--1003},
  year={2022},
  publisher={Nature Publishing Group UK London}
}

@inproceedings{li2024advancing,
  title={Advancing Multi-grained Alignment for Contrastive Language-Audio Pre-training},
  author={Li, Yiming and Guo, Zhifang and Wang, Xiangdong and Liu, Hong},
  booktitle={Proceedings of the 32nd ACM International Conference on Multimedia},
  pages={7356--7365},
  year={2024}
}

@inproceedings{li2026finelap,
  title={FineLAP: Taming Heterogeneous Supervision for Fine-grained Language-Audio Pretraining},
  author={Li, Xiquan and Xu, Xuenan and Ma, Ziyang and Chen, Wenxi and He, Haolin and Kong, Qiuqiang and Chen, Xie},
  booktitle={Proceedings of the 64th Annual Meeting of the Association for Computational Linguistics (Volume 1: Long Papers)},
  pages={10393--10408},
  year={2026}
}

\end{document}